\documentclass[aps,pre,onecolumn,showpacs,showkeys,a4paper]{revtex4}

\usepackage{amsmath}		%outils mathématiques pour latex
\usepackage{amstext}		%insertion de texte dans une formule mathématique
\usepackage{amssymb}		%symboles mathématiques
\usepackage{amsfonts}		%symboles mathématiques
\usepackage{graphicx}		%insertion de graphiques
\usepackage{hyperref}		%références hypertexte

\newcommand{\C}[2]{\left(\begin{array}{c}#1\\#2\end{array}\right)}
\renewcommand{\O}[1]{\mathcal{O}\left(#1\right)}

\begin{document}

\title{Fluctuations and skewness of the current in the partially asymmetric exclusion process}
\author{Sylvain Prolhac}
\affiliation{Institut de Physique Th\'eorique,\\
CEA, IPhT, F-91191 Gif-sur-Yvette, France\\
CNRS, URA 2306, F-91191 Gif-sur-Yvette, France}
\date{May 27, 2008}

\begin{abstract}
We use functional Bethe Ansatz equations to calculate the cumulants of the total current in the partially asymmetric exclusion process. We recover known formulas for the first two cumulants (mean value of the current and diffusion constant) and obtain an explicit finite size formula for the third cumulant. The expression for the third cumulant takes a simple integral form in the limit where the asymmetry scales as the inverse of the square root of the size of the system, which corresponds to a natural separation between weak and strong asymmetry.
\end{abstract}

\maketitle

\begin{section}{Introduction}
The asymmetric simple exclusion process (ASEP) is one of the most simple examples of a stochastic interacting particles model with an out of equilibrium stationary state. It features classical particles hoping on a lattice and interacting through hard core exclusion. It can be seen as a growth model with deposition and evaporation of particles and is thus a discrete version of a system evolving by the Kardar-Parisi-Zhang (KPZ) equation. The one dimensional model has been much studied in the past \cite{L85.1,S91.1,D98.1,S01.1,GM06.1}. It is known to be exactly solvable through various methods including the Bethe Ansatz \cite{D87.1,GS92.1}, which uses the underlying integrability of the Markov matrix of the model (similar to the Hamiltonian of an XXZ spin chain with twisted boundary conditions), and the matrix product Ansatz method \cite{DEHP93.1,BE07.1}, which consists in writing the stationary state eigenvector as a trace of product of matrices.\\\indent
The fluctuations of the current have received much attention \cite{DL98.1,DE99.1,BD05.1}. In the long time limit, the system reaches a stationary state, independent of the initial configuration for finite size systems. A manifestation of the fact the system is not at equilibrium in the stationary state is the existence of a current of particles flowing through the system. In this stationary state, the mean value $J$ of the current is simply proportional to the asymmetry of the rates at which the particles hop to the right and to the left, which can be thought as a driving field. The fluctuations of the current describe how the current differs from its mean value, that is the probability of finding a current $j$ different from $J$. But finding the whole probability density function can be difficult. Instead, one can study its first cumulants. The second cumulant, related to the diffusion constant, describes the width of the probability density whereas the third cumulant represents the asymmetry and the non-gaussianity of the probability density. The fourth cumulant describes how much the peak of the distribution is sharp.\\\indent
The fluctuations of the current have been studied for different variants of the ASEP: finite size open lattice \cite{BD05.1}, systems with second class particles \cite{DE99.1,BFMM02.1}, infinite lattice \cite{PS02.1,S07.1},... For the totally asymmetric exclusion process, in which the particles only hop in one direction (TASEP), on a finite size lattice with periodic boundary conditions, all the cumulants of the probability distribution of the current have been calculated by Bethe Ansatz, using a simplification of the Bethe Ansatz equations that occur only in this case. For the partially asymmetric exclusion process on a ring, a particular scaling limit of the probability distribution function of the current was calculated using a thermodynamic limit of the Bethe Ansatz equations \cite{LK99.1}. However, only the first two cumulants were known for finite size systems \cite{DM97.1,PM08.1}. In the present paper we introduce a method that allows us to calculate exact expressions for the first cumulants. We use it to get the diffusion constant with very few calculations, and obtain an explicit formula (\ref{cumulant3}) for the third cumulant. We also derive a rather simple integral formula (\ref{cumulant3 integral}) for the third cumulant in the scaling limit where the asymmetry goes to zero as the inverse of the square root of the size of the system. This formula is similar to the one obtained in \cite{DM97.1} for the diffusion constant in the same scaling limit.\\\indent
The structure of the paper is as follows: In section \ref{section BA}, we recall the functional Bethe Ansatz equations used in \cite{PM08.1} to calculate the two first cumulants of the total current. In section \ref{section solution BA} we show how to solve perturbatively these equations, giving a way to calculate explicitly the cumulants one after the other. We obtain easily the two first cumulants (mean value of the current and diffusion constant). For the diffusion constant, our derivation is much simpler than in \cite{DM97.1} and \cite{PM08.1}. In section \ref{section cumulant3}, after a more involved calculation, we obtain an explicit expression for the third cumulant. We show that for the cases of either the symmetric exclusion process (SSEP, the particle hop on both sides with the same rate) or TASEP it reduces to known formulas. We finally write the third cumulant formula in an integral form in the scaling limit where the asymmetry goes to zero as the inverse of the square root of the size of the system.
\end{section}

\begin{section}{Bethe Ansatz for the fluctuations of the current}
\label{section BA}
We consider in this paper the partially asymmetric simple exclusion process (PASEP) on a one dimensional lattice with periodic boundary conditions. Each of the $L$ sites can be empty or occupied by at most one of the $n$ particles. The configuration space of the model has then dimension $\Omega=\C{L}{n}$. On this configuration space, we define the following Markov process: each particle can hop to the right with rate $p$ (that is, with probability $pdt$ for an infinitesimal time interval $dt$) and to the left with rate $q$ provided that the target site is empty. Otherwise, the particle cannot move. This dynamics can be encoded in the Markov matrix $M=M_{0}+pM_{1}+qM_{-1}$, whose diagonal part $M_{0}$ contains the exit rates from each configuration, and $M_{1}$ (resp. $M_{-1}$) the incoming rates obtained by moving one particle to the right (resp. to the left).\\\indent
Let $Y_{t}$ be the total distance covered by all the particles between time 0 and $t$. $Y_{t}$ is thus the integrated current between time 0 and $t$. It has been shown \cite{LS99.1,DL98.1,PM08.1} that the generating function of the cumulants of $Y_{t}$ in the long time limit can be obtained from the diagonalization of a deformation of the Markov matrix of the system: $M(\gamma)=M_{0}+pe^{\gamma}M_{1}+qe^{-\gamma}M_{-1}$, introducing the parameter $\gamma$ which can be seen as a fugacity associated with the leaps of the particles. More precisely, the eigenvalue $E(\gamma)$ of $M(\gamma)$ corresponding to the stationary state of the system, that is its eigenvalue with maximal real part, is given by
\begin{equation}
\label{E[cumulants]}
E(\gamma)=\lim_{t\to\infty}\frac{\log\langle e^{\gamma Y_{t}}\rangle}{t}=J\gamma+\frac{\Delta}{2}\gamma^{2}+\frac{E_{3}}{6}\gamma^{3}+\cdots
\end{equation}
$J$ is the total current and $\Delta$ is the diffusion constant.
\begin{align}
&J=\lim_{t\to\infty}\frac{\langle Y_{t}\rangle}{t}\\
&\Delta=\lim_{t\to\infty}\frac{\langle Y_{t}^{2}\rangle-\langle Y_{t}\rangle^{2}}{t}\\
&E_{3}=\lim_{t\to\infty}\frac{\langle Y_{t}^{3}\rangle-3\langle Y_{t}\rangle\langle Y_{t}^{2}\rangle+2\langle Y_{t}\rangle^{3}}{t}
\end{align}
The diagonalization of $M(\gamma)$ can be performed using the Bethe Ansatz, because of the underlying integrability of the model. We are now going to recall the functional equations derived in \cite{PM08.1} for the calculation of $E(\gamma)$. Defining the asymmetry parameter $x$ by
\begin{equation}
x=\frac{q}{p}
\end{equation}
the Bethe equations of the system are
\begin{equation}
\label{eby}
e^{L\gamma}\left(\frac{1-y_{i}}{1-xy_{i}}\right)^L+\prod_{j=1}^n\frac{y_{i}-xy_{j}}{xy_{i}-y_{j}}=0.
\end{equation}
The Bethe roots $y_{i}$ $1\leq i\leq n$ depend on both $x$ and $\gamma$. They can be used to write the corresponding eigenvector as a sum over the $n!$ permutations of the particles. Introducing the polynomial
\begin{equation}
\label{Q[y]}
Q(t)=\prod_{i=1}^{n}(t-y_{i})
\end{equation}
whose zeros are the Bethe roots, it can be shown \cite{PM08.1} that the Bethe equations (\ref{eby}) can be rewritten as functional equations
\begin{equation}
\label{ebQR}
Q(t)R(t)=e^{L\gamma}(1-t)^{L}Q(xt)+(1-xt)^{L}x^{n}Q(t/x),
\end{equation}
where $Q$ is a polynomial of degree $n$ (with a coefficient of highest degree equal to 1) and $R$ is a polynomial of degree $L$. Both $Q$ and $R$ must be determined by solving equation (\ref{ebQR}). This equation is known as Baxter's scalar $TQ$ equation \cite{B82.1,S01.2,N03.1,DDT07.1}. Equation (\ref{ebQR}) sets the value of $R(0)$
\begin{equation}
\label{R(0)}
R(0)=x^{n}+e^{L\gamma}
\end{equation}
and the behavior of $R(t)$ when $t\to\infty$
\begin{equation}
\label{R(infinity)}
R(t)\sim(x^{L}+x^{n}e^{L\gamma})(-1)^{L}t^{L}\text{ when $t\to\infty$}.
\end{equation}
When $\gamma=0$, the solution of the Bethe equations (\ref{eby}) corresponding to the stationary state is $y_{i}=0$ for all $i$. Equivalently, the corresponding solution of (\ref{ebQR}) is $Q(t)=t^{n}$ and $R(t)=x^{n}(1-t)^{L}+(1-xt)^{L}$. Expanding $Q(t)$ and $R(t)$ near $\gamma=0$, we get
\begin{equation}
\label{Qst}
Q(t)=t^{n}+\O{\gamma}
\end{equation}
and
\begin{equation}
\label{Rst}
R(t)=x^{n}(1-t)^{L}+(1-xt)^{L}+\O{\gamma}.
\end{equation}
The stationary state has another interesting property: it is a state with zero momentum, which leads to \cite{PM08.1}
\begin{equation}
\label{Q(1)}
x^{n}Q(1/x)=e^{n\gamma}Q(1)
\end{equation}
or, in terms of the polynomial $R$
\begin{equation}
\label{R(1)}
R(1)=e^{n\gamma}(1-x)^{L}.
\end{equation}
The latter equation does not provide additional information beside (\ref{Rst}), but it is useful to make some calculations easier.\\\indent
Finally, with $Q$ and $R$ solution of (\ref{ebQR}), the eigenvalue of $M(\gamma)$ corresponding to the stationary state is given by \cite{PM08.1}
\begin{equation}
\label{vpQR}
\frac{E(\gamma)}{p}=(1-x)\left(\frac{Q'(1)}{Q(1)}-\frac{1}{x}\frac{Q'(1/x)}{Q(1/x)}\right)=-Lx-(1-x)\frac{R'(1)}{R(1)}.
\end{equation}
Equation (\ref{ebQR}) could be solved perturbatively near $\gamma=0$ right now. Inserting the solution into (\ref{vpQR}), it would give us the first cumulants of the current. This was done in \cite{PM08.1} up to order 2. Here, we will rewrite the previous equations in a different way before making the perturbative expansion near $\gamma=0$. It will make the calculations much simpler than in \cite{PM08.1}, and allow us to calculate the third cumulant. We divide both sides of the functional equation (\ref{ebQR}) by $(1-t)^{L}(1-xt)^{L}Q(t)$
\begin{equation}
\label{R[A]}
\frac{R(t)}{(1-t)^{L}(1-xt)^{L}}=\frac{A(t)}{(1-t)^{L}}+\frac{1}{(1-xt)^{L}}\frac{x^{n}e^{L\gamma}}{A(xt)},
\end{equation}
where we defined
\begin{equation}
\label{A[Q]}
A(t)=x^{n}\frac{Q(t/x)}{Q(t)}.
\end{equation}
From this definition of $A(t)$, and using (\ref{Q(1)}) and the fact that $Q$ is of degree $n$, we know the value of $A(t)$ in $t=0$, $t=1$ and in the limit $t\to\infty$
\begin{equation}
\label{A(0)A(1)A(infinity)}
A(0)=x^{n}\text{,}\qquad A(1)=e^{n\gamma}\qquad\text{and}\qquad\lim_{t\to\infty}A(t)=1
\end{equation}
From (\ref{Qst}), we also know that
\begin{equation}
\label{Ast}
A(t)=1+\O{\gamma}.
\end{equation}
In terms of $A(t)$, the expression (\ref{vpQR}) for the eigenvalue rewrites, using the value of $A(1)$ (\ref{A(0)A(1)A(infinity)})
\begin{equation}
\label{vpA}
\frac{E(\gamma)}{p}=-(1-x)e^{-n\gamma}A'(1).
\end{equation}
\end{section}

\begin{section}{Perturbative solution of the functional equation}
\label{section solution BA}
In this section, we will first show how to eliminate $R(t)$ from equation (\ref{R[A]}), leaving us with a closed equation for $A(t)$. We will see how we can solve this equation perturbatively near $\gamma=0$. Then, we will reformulate this solution to make the explicit calculation of the cumulants easier.

\begin{subsection}{Perturbative solution for $A(t)$}
\label{Section recurr}
From (\ref{Ast}), we can write the expansion of $A(t)$ near $\gamma=0$ as
\begin{equation}
A(t)=1+\sum_{k=1}^{\infty}A_{k}(t)\gamma^{k}.
\end{equation}
But $Q(t)$ is normalized such that $Q(t)-t^{n}$ is a polynomial of degree $n-1$. Thus, from (\ref{A[Q]}), the $A_{k}(t)$ are polynomials in $1/t$ of degree $kn$ without constant term ($A_{k}(t)\to 0$ when $t\to\infty$).\\\indent
From now on, we will look at a perturbative solution of (\ref{R[A]}) for small $\gamma$. $R(t)$, which is a polynomial in $t$, can be seen as a formal series in $t$ and $\gamma$ with only nonnegative powers in $t$. The l.h.s. of equation (\ref{R[A]}) is then also a formal series in $t$ and $\gamma$ with only nonnegative powers in $t$. But we also know that, at each (nonzero) order in $\gamma$, $A(t)$ has only negative powers in $t$ so that the r.h.s. of equation (\ref{R[A]}) has both negative and nonnegative powers in $t$. Thus, equation (\ref{R[A]}) simply means that the negative powers in $t$ in equation (\ref{R[A]}) cancel. We will write this as
\begin{equation}
\label{eqA}
\frac{A(t)}{(1-t)^{L}}+\frac{1}{(1-xt)^{L}}\frac{x^{n}e^{L\gamma}}{A(xt)}=\O{t^{0}},
\end{equation}
which will mean: at each order in $\gamma$, the l.h.s. of (\ref{eqA}) is finite when $t\to 0$. We have eliminated $R(t)$ from equation (\ref{R[A]}). We will see that equation (\ref{eqA}) still determines $A(t)$ uniquely.\\\indent
From now on, every expansion in powers of $t$ will have to be understood as an expansion of a formal series in powers of $\gamma$ followed, at each order in $\gamma$, by an expansion in powers of $t$. In the following, we will use the notation $[f(t)]_{(k)}\equiv[f]_{(k)}$ to refer to the coefficient of the term $t^{k}$ in the expansion of the formal series $f(t)$. We also introduce the notation $[f(t)]_{(-)}$ which will represent the negative powers of $f$ in $t$ (that is terms $t^{k}$ with $k<0$). On the contrary, $[f(t)]_{(+)}$ will represent the nonnegative powers of $f$ in $t$ (terms $t^{k}$ with $k\geq 0$).\\\indent
To solve equation (\ref{eqA}), we will write it in the slightly more complicated form
\begin{equation}
\label{eqA2}
\Delta_{x}\left(\frac{A(t)}{(1-t)^{L}}\right)=-\frac{x^{n}}{(1-xt)^{L}}\left(A(xt)+\frac{e^{L\gamma}}{A(xt)}\right)+\O{t^{0}},
\end{equation}
$\Delta_{x}$ being the operator which acts on an arbitrary function $u(t)=\sum\limits_{k\in\mathbb{Z}}[u]_{(k)}t^{k}$ as
\begin{equation}
\label{Deltax}
(\Delta_{x}u)(t)=u(t)-x^{n}u(xt)=\sum_{k\in\mathbb{Z}}\left(1-x^{n+k}\right)[u]_{(k)}t^{k}.
\end{equation}
As $\Delta_{x}$ gives $0$ when applied to $t^{-n}$, the equation
\begin{equation}
\label{Deltau=v}
(\Delta_{x}u)(t)=v(t)
\end{equation}
can be formally solved as
\begin{equation}
\label{invDeltax}
u(t)=(\Delta_{x}^{-1}v)(t)=\sum_{k\neq -n}\frac{[v]_{(k)}t^{k}}{1-x^{n+k}}-\frac{b}{t^{n}}
\end{equation}
for any formal series $v$ which has no term $t^{-n}$. $b$ is an arbitrary coefficient which is not constrained by equation (\ref{Deltau=v}). We can use this to invert $\Delta_{x}$ in (\ref{eqA2}). Recalling that $A(t)\to 1$ when $t\to\infty$ and that $A(t)-1$ has only negative powers in $t$, we obtain
\begin{equation}
\label{Arecur}
A(t)=1-[(1-t)^{L}\tilde{g}(t)]_{(-)}-b\left[\frac{(1-t)^{L}}{t^{n}}\right]_{(-)},
\end{equation}
with
\begin{equation}
\label{g[A]}
\tilde{g}(t)=\left[\Delta_{x}^{-1}\left(\frac{x^{n}}{(1-xt)^{L}}\left(A(xt)+\frac{e^{L\gamma}}{A(xt)}\right)\right)\right]_{(-)}.
\end{equation}
The term containing $b$ is not constrained by equation (\ref{eqA2}). The condition necessary to invert $\Delta_{x}$ in equation (\ref{eqA2}), which is that the r.h.s. of (\ref{eqA2}) must not contain a term $t^{-n}$, implies that $\tilde{g}(t)$ does not contain a term $t^{-n}$.\\\indent
Written like that, equation (\ref{Arecur}) contains $A$ on both sides (through $\tilde{g}$ on the r.h.s.). However, equation (\ref{Arecur}) gives a recursive solution for $A(t)$ order by order in powers of $\gamma$. Indeed, if we take the term of order $\gamma^{k}$ in this equation, the l.h.s. depends only on $A_{k}(t)$ whereas the r.h.s. depends only on the $A_{j}(t)$ with $j<k$: the $A_{k}(t)$ cancels out at order $k$ in the expansion in powers of $\gamma$ of
\begin{equation}
A(xt)+\frac{e^{L\gamma}}{A(xt)}=1+\sum_{k=1}^{\infty}A_{k}(xt)\gamma^{k}+\frac{1+(e^{L\gamma}-1)}{1+\sum\limits_{k=1}^{\infty}A_{k}(xt)\gamma^{k}}
\end{equation}
Thus, equation (\ref{Arecur}) allows to recursively compute all the orders in $\gamma$ starting from (\ref{Ast}). The parameter $b$ in (\ref{Arecur}) can be set from the value (\ref{A(0)A(1)A(infinity)}) of $A(1)$.\\\indent
Using equations (\ref{Arecur}), (\ref{Ast}) and (\ref{A(0)A(1)A(infinity)}), we can do the calculation for the first cumulants. But we will first turn these equations into another form which will make the calculations a little simpler.\\\indent
We must emphasize that our method involves only algebraic manipulations of formal series in two parameters, namely $\gamma$ and $t$. The Bethe roots $y_{i}(\gamma)$ (which are the zeros of the polynomial $Q(t)$ and the poles of the rational function $A(t)$) have completely disappeared since we did the $\gamma$ expansion first: the $A_{k}(t)$ have only a pole of order $kn$ in $t=0$. We do not have to follow the $y_{i}(\gamma)$ as a function of $\gamma$: we only use some algebraic properties of $A(t)$.
\end{subsection}

\begin{subsection}{A simpler formulation for the perturbative solution}
We will now eliminate $A(t)$ from our recursive equations, and work on the auxiliary quantity $\tilde{g}(t)$. Noticing that $A^{-1}(xt)-1$ has also only negative powers in $t$, we find from equations (\ref{eqA}) and (\ref{Arecur}) that $x^{n}e^{L\gamma}A^{-1}(xt)$ can be written in terms of $\tilde{g}(t)$ as
\begin{equation}
x^{n}e^{L\gamma}A^{-1}(xt)=x^{n}e^{L\gamma}+[(1-xt)^{L}\tilde{g}(t)]_{(-)}+b\left[\frac{(1-xt)^{L}}{t^{n}}\right]_{(-)}.
\end{equation}
We can absorb the $b$ terms of $A(t)$ and $x^{n}e^{L\gamma}A^{-1}(xt)$ into $\tilde{g}$, defining
\begin{equation}
g(t)=\tilde{g}(t)+\frac{b}{t^{n}}.
\end{equation}
$b$ is thus the coefficient of $t^{-n}$ in $g(t)$ as $\tilde{g}(t)$ has no term $t^{-n}$. In the following, we will both need using $g(t)$ and $\tilde{g}(t)$. The expressions for $A(t)$ and $x^{n}e^{L\gamma}A^{-1}(xt)$ become
\begin{align}
\label{A[g]}
A(t)&=1-[(1-t)^{L}g(t)]_{(-)}\\
\label{1/A[g]}
x^{n}e^{L\gamma}A^{-1}(xt)&=x^{n}e^{L\gamma}+[(1-xt)^{L}g(t)]_{(-)}.
\end{align}
From (\ref{g[A]}), at order $k$ in $\gamma$ $g(t)$ is a polynomial in $1/t$ of degree $kn$ without constant term, as are the $A_{k}(t)$. We also see that $g(t)$ is equal to 0 when $\gamma=0$ (as $A(t)=1$ when $\gamma=0$), giving the expansion
\begin{equation}
g(t)=\sum_{k=1}^{\infty}g_{k}(t)\gamma^{k}.
\end{equation}
To find a closed equation for $g(t)$, we eliminate $A(t)$ between (\ref{A[g]}) and (\ref{1/A[g]}), using $(A(t))(x^{n}e^{L\gamma}A^{-1}(t))=x^{n}e^{L\gamma}$:
\begin{equation}
[(1-t)^{L}g(t/x)]_{(-)}-x^{n}e^{L\gamma}[(1-t)^{L}g(t)]_{(-)}-[(1-t)^{L}g(t)]_{(-)}[(1-t)^{L}g(t/x)]_{(-)}=0,
\end{equation}
which can be rewritten as
\begin{equation}
g(t/x)-x^{n}e^{L\gamma}g(t)=\frac{[(1-t)^{L}g(t)]_{(-)}[(1-t)^{L}g(t/x)]_{(-)}}{(1-t)^{L}}+\O{t^{0}}.
\end{equation}
As $g(t)$ is of order $\gamma$, this allows us to solve order by order in powers of $\gamma$. We can simplify the r.h.s. of the previous equation, writing
\begin{align}
&\frac{[(1-t)^{L}g(t)]_{(-)}[(1-t)^{L}g(t/x)]_{(-)}}{(1-t)^{L}}\\
&=\frac{\left((1-t)^{L}g(t)-[(1-t)^{L}g(t)]_{(+)}\right)\left((1-t)^{L}g(t/x)-[(1-t)^{L}g(t/x)]_{(+)}\right)}{(1-t)^{L}}\nonumber\\
&=(1-t)^{L}g(t)g(t/x)-g(t/x)[(1-t)^{L}g(t)]_{(+)}-g(t)[(1-t)^{L}g(t/x)]_{(+)}+\O{t^{0}}\nonumber\\
&=-(1-t)^{L}g(t)g(t/x)+g(t/x)[(1-t)^{L}g(t)]_{(-)}+g(t)[(1-t)^{L}g(t/x)]_{(-)}+\O{t^{0}}\nonumber
\end{align}
and we obtain
\begin{equation}
g(t/x)-x^{n}e^{L\gamma}g(t)=-(1-t)^{L}g(t)g(t/x)+g(t)[(1-t)^{L}g(t/x)]_{(-)}+g(t/x)[(1-t)^{L}g(t)]_{(-)}+\O{t^{0}}
\end{equation}
or, at order $r$ in $\gamma$, multiplying $t$ by $x$
\begin{align}
\label{recurrgainverser}
(\Delta_{x}g_{r})&(t)=x^{n}\sum_{k=1}^{r-1}\frac{L^{k}}{k!}g_{r-k}(xt)\\
&-\sum_{k=1}^{r-1}\left((1-xt)^{L}g_{r-k}(xt)g_{k}(t)-g_{k}(t)[(1-xt)^{L}g_{r-k}(xt)]_{(-)}-g_{r-k}(xt)[(1-xt)^{L}g_{k}(t)]_{(-)}\right)+\O{t^{0}}\nonumber,
\end{align}
where we used once again the operator $\Delta_{x}$ (\ref{Deltax}). Using the formal inversion formula (\ref{invDeltax}) for $\Delta_{x}$, we see that the previous equation for $g(t)$ can be solved if
\begin{equation}
\label{conditionrecurrg}
\sum_{k=1}^{r-1}\left[x^{n}\frac{L^{k}}{k!}g_{r-k}(xt)-(1-xt)^{L}g_{r-k}(xt)g_{k}(t)+g_{k}(t)[(1-xt)^{L}g_{r-k}(xt)]_{(-)}+g_{r-k}(xt)[(1-xt)^{L}g_{k}(t)]_{(-)}\right]_{(-n)}=0,
\end{equation}
which is a consequence of the fact that equation (\ref{recurrgainverser}) holds: it must be true if (\ref{ebQR}) has a solution verifying (\ref{Qst}) and (\ref{Rst}). We obtain
\begin{align}
\label{recurrg}
&\tilde{g}_{r}(t)=\sum_{k=1}^{r-1}\sum_{\substack{l=1\\(l\neq n)}}^{(r-k)n}\frac{x^{n}t^{-l}}{1-x^{n-l}}\left[\frac{L^{k}}{k!}g_{r-k}(xt)\right]_{(-l)}\\
&-\sum_{k=1}^{r-1}\sum_{\substack{l=1\\(l\neq n)}}^{rn}\frac{\left[(1-xt)^{L}g_{r-k}(xt)g_{k}(t)-g_{k}(t)[(1-xt)^{L}g_{r-k}(xt)]_{(-)}-g_{r-k}(xt)[(1-xt)^{L}g_{k}(t)]_{(-)}\right]_{(-l)}}{(1-x^{n-l})t^{l}}\nonumber.
\end{align}
and
\begin{equation}
g_{r}(t)=\frac{b_{r}}{t^{n}}+\tilde{g}_{r}(t)
\end{equation}
The coefficient $b_{r}$ is the term of order $\gamma^{r}$ in $b$. Again, the value (\ref{A(0)A(1)A(infinity)}) of $A(1)$ sets $b_{r}$. We note that this formula does not require to divide by a formal series in $\gamma$, unlike (\ref{Arecur}) where we had to divide by $A(xt)$. The nonlinearity of the recurrence formula is then simpler here; it consists only in a product of two series in $\gamma$.
\end{subsection}
\end{section}

\begin{section}{Explicit calculations for the two first cumulants of the current}
In this section, after expressing the generating function of the cumulants of the current in terms of the perturbative solution described in the previous section, we will calculate explicitly the two first cumulants.

\begin{subsection}{Expression of the stationary state eigenvalue}
We will now express the stationary state eigenvalue using the function $g(t)$. As $g(t)$ has only negative powers in $t$, expression (\ref{A[g]}) for $A(t)$ in terms of $g(t)$ can be written as
\begin{equation}
A(t)=1-\sum_{i=1}^{\infty}\sum_{j=0}^{i-1}[g]_{(-i)}\C{L}{j}(-1)^{j}t^{j-i}.
\end{equation}
Using the binomial coefficients formulas (\ref{Binomial C}) and (\ref{Binomial kC}), we get for the eigenvalue
\begin{equation}
\label{vpg}
\frac{E(\gamma)}{p}=\frac{1-x}{e^{n\gamma}}\sum_{j=1}^{L}{\C{L}{j}(-1)^{j}\frac{j(L-j)}{L(L-1)}}[g]_{(-j)}.
\end{equation}
This expression for $E(\gamma)$ is of order $\gamma$, as is $g(t)$, which means $E(\gamma=0)=0$, as expected for the stationary state.\\\indent
The last equation allows us to calculate the eigenvalue at any order in $\gamma$ if we know $g(t)$ at the corresponding order. $g(t)$ is obtained by the recurrence equation (\ref{recurrg}) in which $b_{r}$ is set using the known value (\ref{A(0)A(1)A(infinity)}) of $A(1)$. Thus, in the calculation of the eigenvalue $b_{r}$ can be eliminated from (\ref{vpg}) where it appears through $g(t)=\tilde{g}(t)+\frac{b}{t^{n}}$. If we want to calculate the eigenvalue at order $3$, we will not need $b_{3}$ (but we will still need $b_{1}$ and $b_{2}$ as the recurrence equation (\ref{recurrg}) involves $g(t)$ and not only $\tilde{g}(t)$). Using the value (\ref{A(0)A(1)A(infinity)}) of $A(1)$,
\begin{equation}
\label{valeur en 1}
\sum_{j=1}^{L}{\C{L}{j}(-1)^{j}\frac{j}{L}[g]_{(-j)}}=e^{n\gamma}-1,
\end{equation}
which gives for $b=\sum_{r=1}^{\infty}b_{r}\gamma^r$
\begin{equation}
\label{b[g]}
\frac{b}{\nu}=\frac{e^{n\gamma}-1}{n}-\sum_{j=1}^{L}{\C{L}{j}(-1)^{j}\frac{j}{nL}[\tilde{g}]_{(-j)}},
\end{equation}
with
\begin{equation}
\nu\equiv(-1)^{n}\frac{L}{\C{L}{n}}
\end{equation}
We finally get for the eigenvalue, putting (\ref{vpg}) and (\ref{b[g]}) together
\begin{equation}
\label{vpgtilde}
\frac{E(\gamma)}{p}=\frac{1-x}{e^{n\gamma}}\left(\sum_{j=1}^{L}{\C{L}{j}(-1)^{j}\frac{j(n-j)}{L(L-1)}}[\tilde{g}]_{(-j)}+\frac{L-n}{L-1}(e^{n\gamma}-1)\right).
\end{equation}
We will now use (\ref{recurrg}), (\ref{b[g]}) and (\ref{vpgtilde}) to calculate $E(\gamma)$ up to order 2 in $\gamma$, recovering known results for the mean value of the current and for the diffusion constant.
\end{subsection}

\begin{subsection}{Calculation of the mean value of the current}
From (\ref{E[cumulants]}), the mean value $J$ of the current is given by
\begin{equation}
\frac{J}{p}=\left(\frac{dE(\gamma)}{d\gamma}\right)_{|\gamma=0}.
\end{equation}
We only need $E(\gamma)$ at the first order in $\gamma$. In this case, the two complicated terms with the sums in (\ref{recurrg}) do not contribute and $\tilde{g}_{1}(t)=0$. Using (\ref{vpgtilde}), the eigenvalue reads
\begin{equation}
\frac{J}{p}=(1-x)\frac{n(L-n)}{L-1}.
\end{equation}
We did not need $b_{1}$ to calculate $J$, but we will need it for the next orders. Using (\ref{b[g]}) we find $b_{1}=\nu$ and
\begin{equation}
\label{g1}
g_{1}(t)=\frac{\nu}{t^{n}}.
\end{equation}
\end{subsection}

\begin{subsection}{Calculation of the diffusion constant}
From (\ref{E[cumulants]}), the diffusion constant $\Delta$ is given by
\begin{equation}
\frac{\Delta}{p}=\left(\frac{d^{2}E(\gamma)}{d\gamma^{2}}\right)_{|\gamma=0}.
\end{equation}
The recurrence (\ref{recurrg}) gives
\begin{equation}
\tilde{g}_{2}(t)=\nu^{2}\sum_{k=1}^{n-1}{\frac{\C{L}{2n-k}(-t)^{-k}}{1-x^{k-n}}}-\nu^{2}\sum_{k=n+1}^{2n}{\frac{\C{L}{2n-k}(-t)^{-k}}{1-x^{k-n}}}.
\end{equation}
The eigenvalue is, using (\ref{vpgtilde}) and (\ref{Binomial kCC})
\begin{equation}
\label{cumulant2}
\frac{\Delta}{p}=2(1-x)\frac{L}{L-1}\sum_{k>0}{k^{2}\frac{\C{L}{n+k}\C{L}{n-k}}{\C{L}{n}^{2}}\frac{1+x^{k}}{1-x^{k}}}.
\end{equation}
The derivation of this expression is much simpler than the previous derivation using the Bethe Ansatz \cite{PM08.1}. It uses a little more formalism, but nearly no calculation is needed, contrary to the previous derivation in which many unexpected simplifications occurred in the end of the calculation.\\\indent
The condition (\ref{conditionrecurrg}) necessary for the consistency of equation (\ref{recurrgainverser}) is easily checked
\begin{equation}
Lx^{n}[g_{1}(t)]_{(-n)}-[(1-t)^{L}g_{1}(t)g_{1}(t/x)]_{(-n)}=0.
\end{equation}
As for the first order, we did not need $b_{2}$ to calculate the diffusion constant, but we will need it for the third order. We set $b_{2}$ using equation (\ref{b[g]}) and the binomial formula (\ref{Binomial kCC})
\begin{equation}
\frac{b_{2}}{\nu}=\frac{L}{2}+\frac{\nu^{2}}{L}\sum_{k=1}^{n}{\C{L}{n+k}\C{L}{n-k}\frac{1+x^{k}}{1-x^{k}}}
\end{equation}
and obtain for $g_{2}(t)$
\begin{equation}
\label{g2}
g_{2}(t)=\nu^{2}\sum_{k=1}^{n-1}{\frac{\C{L}{2n-k}(-t)^{-k}}{1-x^{k-n}}}-\nu^{2}\sum_{k=n+1}^{2n}{\frac{\C{L}{2n-k}(-t)^{-k}}{1-x^{k-n}}}+\frac{L\nu}{2t^{n}}+\frac{\nu^{3}}{Lt^{n}}\sum_{k=1}^{n}{\C{L}{n+k}\C{L}{n-k}\frac{1+x^{k}}{1-x^{k}}}.
\end{equation}
\end{subsection}
\end{section}

\begin{section}{Third cumulant of the current}
\label{section cumulant3}
This section is devoted to the third cumulant of the current in the stationary state of PASEP. First, we will explain the steps needed to derive expression (\ref{cumulant3}) for the third cumulant, leaving the complete proof to appendix \ref{appendix ordre3}. We will then show that we recover known formulas in the case of TASEP and SSEP. Finally, we will write the expression for the third cumulant as a double integral in the scaling limit where the asymmetry goes to zero as the inverse of the square root of the size of the system.

\begin{subsection}{Exact formula}
We will now calculate the third cumulant of the current. Without loss of generality, we can suppose that $L\leq 2n$ because of the particle-hole symmetry. It will be easier because all the sums will have their bounds between 0 and $2n$. The steps are the same as for the two first orders: first, calculate $\tilde{g}_{3}(t)$ using the recurrence relation (\ref{recurrg}). Then, insert it into (\ref{vpgtilde}) to get the eigenvalue. This time, as we do not want the next order, we will not need $b_{3}$. The only big difference will be that these two steps use now longer expressions involving double sums (instead of simple sums for the second order and no sum at all for the first order). In the end of the calculation, which is detailed in appendix \ref{appendix ordre3}, we find that the third cumulant $E_{3}$ is given by
\begin{align}
\label{cumulant3}
&\frac{(L-1)E_{3}}{p(1-x)L^{2}}=6\sum_{i>0}\sum_{j>0}\frac{\C{L}{n+i}\C{L}{n-i}\C{L}{n+j}\C{L}{n-j}}{\C{L}{n}^{4}}(i^{2}+j^{2})\frac{1+x^{i}}{1-x^{i}}\frac{1+x^{j}}{1-x^{j}}\\
&-3\sum_{i>0}\sum_{j>0}\frac{\C{L}{n+i}\C{L}{n+j}\C{L}{n-i-j}+\C{L}{n-i}\C{L}{n-j}\C{L}{n+i+j}}{\C{L}{n}^{3}}(i^{2}+ij+j^{2})\frac{1+x^{i}}{1-x^{i}}\frac{1+x^{j}}{1-x^{j}}\nonumber\\
&-3\sum_{i>0}\frac{\C{L}{n+i}\C{L}{n-i}}{\C{L}{n}^{2}}(i^{2})\left(\frac{1+x^{i}}{1-x^{i}}\right)^{2}+\frac{3n(L-n)}{2(2L-1)}\frac{\C{2L}{2n}}{\C{L}{n}^{2}}-\frac{n(L-n)}{(3L-1)}\frac{\C{3L}{3n}}{\C{L}{n}^{3}}\nonumber.
\end{align}
We checked this formula numerically for systems with $2\leq L\leq 12$, $1\leq n\leq L/2$ and $x=0,0.1,...,0.9$. For all these cases, we computed the sequence
\begin{equation}
a_{k}=\frac{1}{\epsilon}\frac{d^{3}}{d\gamma^{3}}\left(\frac{\langle 1|(1+\epsilon M(\gamma))^{k}|1\rangle}{\langle 1|(1+\epsilon M(\gamma))^{k-1}|1\rangle}-1\right)_{|\gamma=0}
\end{equation}
for $1\leq k\leq 200$ and $\epsilon=\frac{1}{(1+x)(n+1)}$, $\gamma$ being kept a formal parameter. Using convergence acceleration techniques (fitting our sequences with a sum of 2 exponentials using Shanks transformation), we found that, for all the systems considered, the relative error on the third cumulant was at most $1.4\;10^{-9}$, which validates formula (\ref{cumulant3}) for the third cumulant.
\end{subsection}

\begin{subsection}{Some special cases}
\begin{subsubsection}{Totally asymmetric exclusion process}
In the case of the totally asymmetric exclusion process ($x=0$, the particles only hop to the right), using the binomial formulas (\ref{Binomial CC}), (\ref{Binomial k2CC}) (twice), (\ref{Binomial ijCCC+}) and (\ref{Binomial (i2+j2)CCC+}), all the sums in the expression (\ref{cumulant3}) for the third cumulant can be calculated, giving the result
\begin{equation}
\frac{E_{3}(x=0)}{p}=\frac{3L^{2}n(L-n)}{(L-1)(2L-1)}\frac{\C{2L}{2n}^{2}}{\C{L}{n}^{4}}-\frac{4L^{2}n(L-n)}{(L-1)(3L-1)}\frac{\C{3L}{3n}}{\C{L}{n}^{3}},
\end{equation}
which is the same as formula (13) in \cite{DL98.1}.
\end{subsubsection}
\begin{subsubsection}{Symmetric exclusion process}
In the case of the symmetric exclusion process ($x=1$, the particles hop to the right and to the left with the same rate), we see that 3 of the terms of formula (\ref{cumulant3}) are singular. But it turns out that all these singularities cancel, as well as all the constants terms, giving $E_{3}=0$. Indeed, we have the expansion near $x=1$
\begin{equation}
\frac{1+x^{i}}{1-x^{i}}\frac{1+x^{j}}{1-x^{j}}=\frac{4}{ij}\frac{1}{(x-1)^{2}}+\frac{4}{ij}\frac{1}{x-1}+\mathcal{O}(1).
\end{equation}
The coefficients of $(x-1)^{-1}$ and of $(x-1)^{-2}$ are the same, so we only have one calculation to do to cancel both the divergent and the constant term of $E_{3}$ when $x\to 1$. Using the binomial formulas (\ref{Binomial kCC}) and (\ref{Binomial CC CCC+}), we find that both terms vanish, leaving us with
\begin{equation}
E_{3}(x=1)=0.
\end{equation}
This result was expected: all the odd cumulants vanish for the symmetric exclusion process as the system is invariant if we exchange the right and the left. The probability density function of the current is even.
\end{subsubsection}
\end{subsection}

\begin{subsection}{Scaling limit when $x\rightarrow 1$ as $L^{-1/2}$}
The behavior of the system strongly depends on whether $x=1$ or not. If $x=1$, the rates at which the particles hop to the left and to the right are equal and the system belongs to the universality class of the Edwards-Wilkinson equation. If $x\neq 1$, the rates are not symmetric anymore and the system belongs to the universality class of the Kardar-Parisi-Zhang equation. The separation between these two regimes is given by the scaling $1-x\sim L^{-1/2}$ with fixed ratio $\rho=n/L$. This scaling is a rather natural separation between weak and strong asymmetry. Indeed, a particle makes $R\propto (1-x)\Delta t/L$ revolutions through the periodic lattice during a time interval $\Delta t$, from the value of the current. The natural time interval to be considered corresponds to the time necessary for the system to reach its stationary state. This time scales as $L^{z}$ where the dynamical exponent $z$ equals $3/2$ for the ASEP \cite{GS92.1}. Thus, $1-x\sim L^{-1/2}$ is the scaling corresponding to a number of rotations $R\sim 1$ during the time necessary for the system to reach its stationary state. If $1-x\ll L^{-1/2}$, the number of rotations during this time is $\ll 1$ and the system behaves like the symmetric exclusion process. On the contrary, if $1-x\gg L^{-1/2}$, the number of rotations is $\gg 1$ and the system behaves like the totally asymmetric exclusion process.\\\indent
In \cite{DM97.1}, the scaling limit $1-x\sim 1/\sqrt{L}$ of the diffusion constant was calculated from the exact expression (\ref{cumulant2}). It was found (up to a factor $L^{2}$ due to the fact that they calculated the current through a bond) that
\begin{equation}
\label{h2}
\frac{\Delta}{p}\sim \Phi h_{2}(\Phi)\rho(1-\rho)L\quad\text{with}\quad h_{2}(\Phi)=4\int_{0}^{\infty}{du\frac{u^{2}}{\tanh(\Phi u)}e^{-u^{2}}}
\end{equation}
and
\begin{align}
\label{x(f)}
&x=e^{-f}\\
\label{Phi(f)}
&\Phi=\frac{f\sqrt{L\rho(1-\rho)}}{2},
\end{align}
$\Phi$ being held constant in this scaling limit. We will do the same here for the third cumulant in the same scaling limit. From the definition (\ref{x(f)}) of $f$, we have
\begin{equation}
\frac{1+x^{i}}{1-x^{i}}=\frac{1}{\tanh\left(\frac{if}{2}\right)}=\frac{1}{\tanh\left(\frac{i\Phi}{\sqrt{L\rho(1-\rho)}}\right)}.
\end{equation}
When $L$ goes to infinity with fixed $\Phi$, $f\to 0$, so that $f\sim 1-x$ and $(1-x)\sqrt{L}$ is also kept constant, which is precisely the wanted scaling limit. Letting $L$ go to infinity with fixed $\rho$ and $\Phi$, the expression (\ref{cumulant3}) of the third cumulant becomes, using Stirling's approximation for the binomial coefficients
\begin{align}
\label{cumulant3 Stirling}
\frac{E_{3}}{2p\Phi}\sqrt{\frac{\rho(1-\rho)}{L}}\sim&6\sum_{i>0}\sum_{j>0}\frac{(i^{2}+j^{2})e^{-\frac{i^{2}+j^{2}}{L\rho(1-\rho)}}}{\tanh\left(\frac{i\Phi}{\sqrt{L\rho(1-\rho)}}\right)\tanh\left(\frac{j\Phi}{\sqrt{L\rho(1-\rho)}}\right)}-6\sum_{i>0}\sum_{j>0}\frac{(i^{2}+ij+j^{2})e^{-\frac{i^{2}+ij+j^{2}}{L\rho(1-\rho)}}}{\tanh\left(\frac{i\Phi}{\sqrt{L\rho(1-\rho)}}\right)\tanh\left(\frac{j\Phi}{\sqrt{L\rho(1-\rho)}}\right)}\nonumber\\
&-3\sum_{i>0}\frac{i^{2}e^{-\frac{i^{2}}{L\rho(1-\rho)}}}{\tanh^{2}\left(\frac{i\Phi}{\sqrt{L\rho(1-\rho)}}\right)}+\frac{3\sqrt{\pi}}{4}L^{3/2}(\rho(1-\rho))^{3/2}-\frac{2\pi}{3\sqrt{3}}\rho^{2}(1-\rho)^{2}L^{2}.
\end{align}
We only wrote the dominant behavior of each term. The sums can now be written as Riemann integrals over the variables $u=i/\sqrt{L\rho(1-\rho)}$ and $v=j/\sqrt{L\rho(1-\rho)}$
\begin{align}
\frac{E_{3}}{p}\sim&12\Phi(\rho(1-\rho))^{3/2}L^{5/2}\int_{0}^{\infty}\!\!\!\int_{0}^{\infty}{dudv\frac{u^{2}+v^{2}-(u^{2}+uv+v^{2})e^{-uv}}{\tanh(\Phi u)\tanh(\Phi v)}e^{-u^{2}-v^{2}}}\\
-&6\Phi\rho(1-\rho)L^{2}\int_{0}^{\infty}{du\frac{u^{2}}{\tanh^{2}{\Phi u}}e^{-u^{2}}}+\frac{3\sqrt{\pi}}{2}\Phi\rho(1-\rho)L^{2}-\frac{4\pi}{3\sqrt{3}}\Phi(\rho(1-\rho))^{3/2}L^{5/2}\nonumber.
\end{align}
When $L\to\infty$, we get
\begin{equation}
\label{cumulant3 integral}
\frac{E_{3}}{p}\sim -\Phi h_{3}(\Phi)(\rho(1-\rho))^{3/2}L^{5/2},
\end{equation}
with
\begin{equation}
\label{h3}
h_{3}(\Phi)=\frac{4\pi}{3\sqrt{3}}-12\int_{0}^{\infty}\!\!\!\int_{0}^{\infty}{dudv\frac{(u^{2}+v^{2})-(u^{2}+uv+v^{2})e^{-uv}}{\tanh(\Phi u)\tanh(\Phi v)}e^{-u^{2}-v^{2}}}.
\end{equation}
\begin{figure}
\includegraphics[scale=0.5]{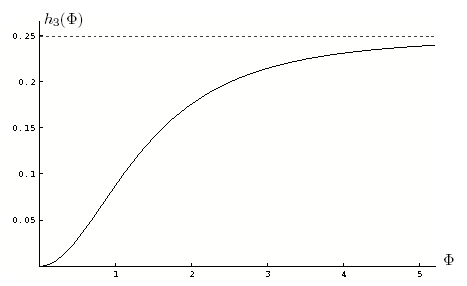}
\caption{Plot of the function $h_{3}(\Phi)$ defined in (\ref{h3}), the asymmetry coefficient $\Phi\sim(1-x)\sqrt{L}$ being defined in (\ref{Phi(f)}). In the limit where the size of the system $L$ goes to infinity and the rates asymmetry $1-x$ goes to $0$ with fixed $\Phi$, $h_{3}(\Phi)$ is proportional to the third cumulant of the current $E_{3}$ divided by $\Phi$. The dashed line indicates the limit of $h_{3}(\Phi)$ when $\Phi$ goes to infinity, whose approximate value is $0.2488186$.}
\label{fig h3}
\end{figure}
A numerical evaluation of $h_{3}(\Phi)$ tells us that it is a monotonic function (see fig. \ref{fig h3}). It grows from $0$ for $\Phi=0$ to $\frac{16\pi}{3\sqrt{3}}-3\pi\approx 0.2488186...$ when $\Phi\rightarrow\infty$. More precisely, the behavior of $h_{3}(\Phi)$ in these limits is
\begin{align}
\label{h3(0)}
&h_{3}(\Phi)\sim\frac{2}{15}\Phi^{2}\qquad\text{when $\Phi\rightarrow 0$}\\
\label{h3(infinity)}
&h_{3}(\Phi)\rightarrow\frac{16\pi}{3\sqrt{3}}-3\pi\qquad\text{when $\Phi\rightarrow\infty$},
\end{align}
as can be seen using
\begin{align}
&\frac{1}{\tanh\Phi u\tanh\Phi v}\to 1\qquad\text{when $\Phi\to\infty$}\\
&\frac{1}{\tanh\Phi u\tanh\Phi v}=\frac{1}{uv\Phi^{2}}+\frac{1}{3}\left(\frac{u}{v}+\frac{v}{u}\right)+\left(-\frac{u^{3}}{45v}+\frac{uv}{9}-\frac{v^{3}}{45u}\right)\Phi^{2}+\O{\Phi^{4}}\qquad\text{when $\Phi\to 0$}
\end{align}
and performing the integrals in polar coordinates.
\begin{subsubsection}{Range of validity of the integral formula for the third cumulant}
So far, we proved the integral formula (\ref{cumulant3 integral}) for the third cumulant in the scaling limit $1-x\sim 1/\sqrt{L}$. We will now see that it holds in fact for more general values of the asymmetry once we extract from it the factor $1-x$ which is in front of the finite size expression (\ref{cumulant3}). More precisely, we will show that
\begin{equation}
\label{cumulant3 scaling}
\frac{E_{3}(x)}{p}\sim -(1-x)\frac{h_{3}(\Phi)}{2}\rho^{2}(1-\rho)^{2}L^{3}\quad\text{for}\quad 1-x\sim 1/L^{r}\text{ with }0\leq r<1,
\end{equation}
$\Phi$ being understood as the function of $x$, $\rho$ and $L$ defined in equations (\ref{x(f)}) and (\ref{Phi(f)}). The case $1-x\gg 1/\sqrt{L}$, which corresponds to taking the limit $\phi\to\infty$, is the easiest. In this limit, all the expressions remain bounded and nonzero, and $h_{3}(\Phi)$ simply converges to its $\Phi\to\infty$ limit (\ref{h3(infinity)}) giving
\begin{equation}
\frac{E_{3}(1-x\gg 1/\sqrt{L})}{p}\sim-(1-x)\left(\frac{8\pi}{3\sqrt{3}}-\frac{3\pi}{2}\right)\rho^{2}(1-\rho)^{2}L^{3}\quad\text{when $L\to\infty$}.
\end{equation}
This is precisely equation (37) of \cite{LK99.1} for the third cumulant when $x\neq 1$. If we take $x=0$, we obtain the TASEP result \cite{DL98.1}.\\\indent
The limit $1-x\ll 1/\sqrt{L}$, which corresponds to $\Phi\to 0$, is more difficult to obtain. From equation (\ref{h3(0)}), we see that in this limit our integral expression for the third cumulant (\ref{cumulant3 integral}) is of order $\Phi^{3}L^{5/2}$. We also know that the third cumulant is equal to $0$ when $x=1$ ($\Phi=0$), so that $E_{3}$ is at least proportional to $\Phi$. However, it is not required to be of order $\Phi^{3}$. Thus, if $\Phi$ is small enough, subdominant terms in $L$ proportional to $\Phi$ may become larger than the $\Phi^{3}L^{5/2}$ term in equation (\ref{cumulant3 Stirling}). In appendix \ref{appendix Phi0}, we show that the correction to equation (\ref{cumulant3 Stirling}) at order $L^{2}$ vanish. We also compute the correction of order $L^{3/2}$ (\ref{cumulant3 subleadingintegral}) and show that it is equal, in the limit $\Phi\to 0$, to $2\Phi\rho^{3/2}(1-\rho)^{3/2}L^{3/2}=(1-x)\rho^{2}(1-\rho)^{2}L^{2}$. This is to be compared with the limit $\Phi\to 0$ of the leading order integral formula (\ref{cumulant3 integral})
\begin{equation}
\label{E3Phi0}
\frac{E_{3}(1-x\ll 1/\sqrt{L})}{p}\sim-\frac{(1-x)^{3}\rho^{3}(1-\rho)^{3}L^{4}}{60}.
\end{equation}
If $1-x\sim 1/L$, these two expressions are of the same order. Thus, when $1-x\sim\sigma/L$ with fixed $\sigma$, the third cumulant is given by
\begin{equation}
\label{E3WASEP}
\frac{E_{3}(1-x\sim\sigma/L)}{p}\sim\sigma\rho^{2}(1-\rho)^{2}L-\frac{\sigma^{3}}{60}\rho^{3}(1-\rho)^{3}L,
\end{equation}
which agrees with \cite{ADLW08.1}. No other subleading correction can become larger when $\Phi\to 0$ as each correction must be at least of order $\Phi$.\\\indent
We finally see that there are two interesting scaling limits for the third cumulant of the current: $1-x\sim 1/\sqrt{L}$ and $1-x\sim 1/L$. Both are in fact natural separations between weak and strong asymmetry. Indeed, if one takes the dynamical exponent to be equal to $2$ (symmetric exclusion process) and not $3/2$ in the discussion we made before equation (\ref{h2}), we find that the natural scaling becomes $1-x\sim 1/L$ and not $1-x\sim 1/\sqrt{L}$. In the scaling limit $1-x\sim 1/\sqrt{L}$, the third cumulant is given by equation (\ref{cumulant3 scaling}) while in the scaling limit $1-x\sim 1/L$ it is given by equation (\ref{E3WASEP}). The first expression is valid as long as $1-x\sim 1/L^{r}$ with $0\leq r<1$. The second expression is valid for $1-x\sim 1/L^{r}$ with $r>1/2$. In the region $1/2<r<1$, both expressions agree.
\end{subsubsection}
\begin{subsubsection}{Normalized third cumulant: skewness}
\begin{figure}
\includegraphics[scale=0.5]{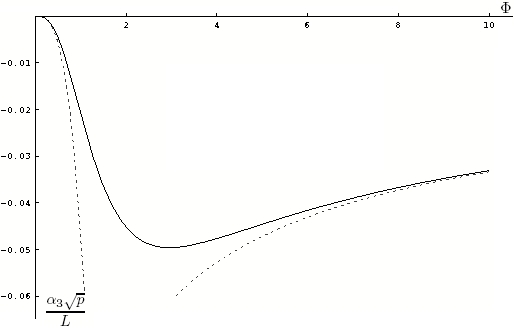}
\caption{Plot of the skewness $\alpha_{3}$ of the probability distribution of the current as a function of the asymmetry in the partially asymmetric exclusion process. The rates for hoping to the right and to the left are $p$ and $xp$ in the scaling limit for which the asymmetry coefficient $\Phi\sim(1-x)\sqrt{L}$ defined in (\ref{Phi(f)}) is held constant when the size of the system goes to infinity. We represented here $\alpha_{3}\sqrt{p}/L$, given in equation (\ref{alpha3}), as a function of $\Phi$. The dashed curves are the asymptotics of the skewness when $\Phi\to 0$ ($\alpha_{3}\sim\Phi^{3}$) and $\Phi\to\infty$ ($\alpha_{3}\sim 1/\sqrt{\Phi}$).}
\label{fig alpha3}
\end{figure}
The skewness of the probability distribution of the current $\alpha_{3}=E_{3}/\Delta^{3/2}$ is given, in the scaling limit $1-x\sim 1/\sqrt{L}$, by (see fig. \ref{fig alpha3})
\begin{equation}
\label{alpha3}
\frac{\alpha_{3}\sqrt{p}}{L}\sim-\frac{h_{3}(\Phi)}{\sqrt{\Phi}(h_{2}(\Phi))^{3/2}}.
\end{equation}
In the limits $\Phi\to 0$ and $\Phi\to\infty$, it behaves as
\begin{align}
&\frac{\alpha_{3}\sqrt{p}}{L}\sim-\frac{\Phi^{3}}{15\sqrt{2}}\qquad\text{when $\Phi\rightarrow 0$}\\
&\frac{\alpha_{3}\sqrt{p}}{L}\sim-\frac{\pi^{1/4}}{\sqrt{\Phi}}\left(\frac{16}{3\sqrt{3}}-3\right)\qquad\text{when $\Phi\rightarrow\infty$},
\end{align}
In particular, $\alpha_{3}\sqrt{p}/L$ goes to $0$ when $\Phi\to 0$, which corresponds to the totally asymmetric case. Indeed for TASEP the diffusion constant $\Delta$ scales as $L^{3/2}$ while the third cumulant scales as $L^{3}$, so that $\alpha_{3}\sim L^{3/4}\ll L$.\\\indent
We see that, at least in the scaling limit for which the asymmetry goes to $0$ as $L^{-1/2}$, the skewness of the total current is negative. This indicates that the left tail of the probability distribution of the total current is larger than its right tail, which means that the current of the system is lower than its mean value with higher probability. This can be understood by a simple argument \cite{DL98.1}: in order to reduce the current of the system, one has to reduce the speed of only one of the particles, as particles can not overtake each other. On the contrary, in order to increase the current of the system, one has to speed up all the particles as the current is limited by the slowest one. It is therefore natural to find a negative skewness for the current.
\end{subsubsection}
\end{subsection}
\end{section}

\begin{section}{Conclusion}
We have presented here an explicit method allowing the calculation of the first cumulants of the total current for PASEP with a finite number of particles in a finite periodic one dimensional lattice. This method, based on the underlying integrability of the model, relies on a perturbative resolution of the functional Bethe Ansatz equation introduced in \cite{PM08.1} to calculate the first two cumulants. We obtained exact formulas for the three first cumulants: in particular, we recovered the known formulas for the first two, but with significantly less calculations, and gave an explicit formula (\ref{cumulant3}) for the third cumulant. This formula is much more complicated than the one for the diffusion constant, involving five terms, two of which are double sums. However, in the limit where the asymmetry $x$ goes to 1 as the inverse of the square root of the size of the system, it takes a simple integral form (\ref{cumulant3 integral}) as does the diffusion constant in the same limit.\\\indent
The perturbative solution formulated here can be used to get all the cumulants of the current for the weakly asymmetric exclusion process, for which the asymmetry scales as the inverse of the size of the system. It gives both the leading and sub-leading terms (work in preparation). It would be interesting to be able to do the same when the asymmetry scales as the inverse of the square root of the size of the system, where nice integral formulas exist for the diffusion constant and the third cumulant. We also think that these methods of functional Bethe Ansatz could be used for other systems for which the Bethe Ansatz equations look quite similar to the ones of PASEP. In particular for open exclusion process, for which the Bethe equations have been found recently \cite{dGE05.1,G08.1}. Or for exclusion process with different classes of particles, for which there has already been work on the fluctuations of the current \cite{DE99.1}.\\\indent
As the asymmetric exclusion process belongs to the universality class of the Kardar-Parisi-Zhang equation, the cumulants of the current for this discrete model in some scaling limit should be characteristic of the KPZ equation. The ratio of the square of the third cumulant with the product of the second and the fourth cumulants was studied numerically by Monte Carlo simulations for TASEP and for two others one dimensional growth models which also belong to the KPZ universality class \cite{DA99.1}. The numerical values obtained there were then compared to the large size limit of the exact result for TASEP derived by Bethe Ansatz \cite{DL98.1}. Something similar could be done for the third cumulant of PASEP we derived here.\\\indent
For a random variable with gaussian probability distribution, only the first and second cumulants are nonzero. The skewness of a probability distribution, defined as the third cumulant, divided by the second to the power $3/2$, is then the first sign of non gaussianity visible on the cumulants. It indicates the asymmetry of the probability distribution. It was recently measured for the distribution of the current in mesoscopic one dimensional systems subject to an electric driving field \cite{R05.1}. It would be interesting to find some real physical system for which the third cumulant of the current is given by our integral formula.

\begin{subsection}*{Acknowledgments}
I thank O. Golinelli and K. Mallick for many useful discussions. I also thank  K. Mallick for his careful reading of the manuscript.
\end{subsection}
\end{section}

\appendix
\begin{section}{Binomial coefficients formulas}
\label{appendix binomial sums}
In the course of this paper, we used the following binomial coefficient formulas
\begin{equation}
\label{Binomial C}
\sum_{k=0}^{n-1}{\C{L}{k}(-1)^{k}}=-(-1)^{n}\frac{n}{L}\C{L}{n}
\end{equation}
\begin{equation}
\label{Binomial kC}
\sum_{k=0}^{n-1}{k\C{L}{k}(-1)^{k}}=-(-1)^{n}n\frac{n-1}{L-1}\C{L}{n}
\end{equation}
\begin{equation}
\label{Binomial CC}
\sum_{k=1}^{n}{\C{L}{n+k}\C{L}{n-k}}=\frac{1}{2}\C{2L}{2n}-\frac{1}{2}\C{L}{n}^{2}
\end{equation}
\begin{equation}
\label{Binomial kCC}
\sum_{k=1}^{n}{k\C{L}{n+k}\C{L}{n-k}}=\frac{n(L-n)}{2L}\C{L}{n}^{2}
\end{equation}
\begin{equation}
\label{Binomial k2CC}
\sum_{k=1}^{n}{k^{2}\C{L}{n+k}\C{L}{n-k}}=\frac{n(L-n)}{2(2L-1)}\C{2L}{2n}
\end{equation}
\begin{align}
\label{Binomial CCC+}
&\sum_{i>0}\sum_{j>0}{\left[\C{L}{n+i}\C{L}{n+j}\C{L}{n-i-j}+\C{L}{n-i}\C{L}{n-j}\C{L}{n+i+j}\right]}\\
&\qquad\qquad\qquad\qquad\qquad\qquad\qquad\qquad\qquad\qquad\qquad\qquad\qquad\qquad=\frac{1}{3}\C{3L}{3n}-\C{L}{n}\C{2L}{2n}+\frac{2}{3}\C{L}{n}^{3}\nonumber
\end{align}
\begin{align}
\label{Binomial (i+j)CCC+}
&\sum_{i>0}\sum_{j>0}{(i+j)\left[\C{L}{n+i}\C{L}{n+j}\C{L}{n-i-j}+\C{L}{n-i}\C{L}{n-j}\C{L}{n+i+j}\right]}\\
&\qquad\qquad\qquad\qquad\qquad\qquad\qquad\qquad\qquad\qquad\qquad\qquad\qquad=\frac{2}{3}\frac{n(L-n)}{L}\C{L}{n}\C{2L}{2n}-\frac{n(L-n)}{L}\C{L}{n}^{3}\nonumber
\end{align}
\begin{align}
\label{Binomial ijCCC+}
&\sum_{i>0}\sum_{j>0}{ij\left[\C{L}{n+i}\C{L}{n+j}\C{L}{n-i-j}+\C{L}{n-i}\C{L}{n-j}\C{L}{n+i+j}\right]}\\
&\qquad\qquad\qquad\qquad\qquad\qquad\qquad\qquad\qquad\qquad\qquad\qquad\qquad=-\frac{1}{3}\frac{n(L-n)}{3L-1}\C{3L}{3n}+\frac{2}{3}\frac{n^{2}(L-n)^{2}}{L^{2}}\C{L}{n}^{3}\nonumber
\end{align}
\begin{align}
\label{Binomial (i2+j2)CCC+}
&\sum_{i>0}\sum_{j>0}{(i^{2}+j^{2})\left[\C{L}{n+i}\C{L}{n+j}\C{L}{n-i-j}+\C{L}{n-i}\C{L}{n-j}\C{L}{n+i+j}\right]}\\
&\qquad\qquad\qquad\qquad\qquad\qquad\qquad=\frac{4}{3}\frac{n(L-n)}{3L-1}\C{3L}{3n}-\frac{n(L-n)}{2L-1}\C{L}{n}\C{2L}{2n}-\frac{2}{3}\frac{n^{2}(L-n)^{2}}{L^{2}}\C{L}{n}^{3}\nonumber
\end{align}
\begin{align}
\label{Binomial (i+j)CCC-}
&\sum_{i>0}\sum_{j>0}{(i+j)\left[\C{L}{n+i}\C{L}{n+j}\C{L}{n-i-j}-\C{L}{n-i}\C{L}{n-j}\C{L}{n+i+j}\right]}\\
&\qquad\qquad\qquad\qquad\qquad\qquad\qquad\qquad\qquad\qquad\qquad\qquad\qquad\qquad\qquad\qquad=-\frac{n(L-n)(L-2n)}{3L^{2}}\C{L}{n}^{3}\nonumber
\end{align}
\begin{align}
\label{Binomial CC CCC+}
&\sum_{i>0}\left(i+\frac{j}{2}\right)\left[\C{L}{n+i}\C{L}{n+j}\C{L}{n-i-j}+\C{L}{n-i}\C{L}{n-j}\C{L}{n+i+j}\right]\\
&\qquad\qquad\qquad\qquad\qquad\qquad\qquad\qquad\qquad\qquad\qquad\qquad\qquad=\left(\frac{n(L-n)}{L}-\frac{j}{2}\right)\C{L}{n}\C{L}{n+j}\C{L}{n-j}\nonumber
\end{align}
\begin{align}
\label{Binomial CC CCC-}
&\sum_{i>0}\left(i+\frac{j}{2}\right)\left[\C{L}{n+i}\C{L}{n+j}\C{L}{n-i-j}-\C{L}{n-i}\C{L}{n-j}\C{L}{n+i+j}\right]\\
&\qquad\qquad\qquad\qquad\qquad\qquad\qquad\qquad\qquad\qquad\qquad\qquad\qquad\qquad=-\frac{j(L-2n)}{2L}\C{L}{n}\C{L}{n+j}\C{L}{n-j}\nonumber
\end{align}
We recall that the binomial coefficient $\C{b}{a}$ is defined for an integer $a$ as
\begin{align}
&\C{b}{a}=\frac{b(b-1)\cdots(b-a+1)}{a!}\qquad\text{if $a>0$}\nonumber\\
&\C{b}{0}=1\qquad\text{for $a=0$}\\
&\C{b}{a}=0\qquad\text{if \quad $a<0$}\nonumber.
\end{align}
In particular, if $b$ is a positive integer
\begin{equation}
\C{b}{a}=0\qquad\text{if $a>b$}.
\end{equation}
Formulas (\ref{Binomial C}) and (\ref{Binomial kC}) can be easily proved using Pascal's triangle formula $\C{b-1}{a-1}+\C{b-1}{a}=\C{b}{a}$ (once for the first equation, twice for the second).\\\indent
We will go on with formulas (\ref{Binomial kCC}), (\ref{Binomial CC CCC+}) and (\ref{Binomial CC CCC-}), which are the easiest. Indeed, if we call $F$ the function of the summation index that we want to sum (the summand), one can find a ``discrete primitive'' $G$ of $F$ with respect to the summation variable, that is, if we call $k$ the summation variable, $F(k)$ can be written as
\begin{equation}
F(k)=G(k+1)-G(k).
\end{equation}
Knowing this expression for $F(k)$, it is then easy to do the summation for any range of summation. The $F$ functions are respectively
\begin{align}
F_{\ref{Binomial kCC}}(k)&=k\C{L}{n+k}\C{L}{n-k}\\
F_{\ref{Binomial CC CCC+}+\ref{Binomial CC CCC-}}(i)&=\left(i+\frac{j}{2}\right)\C{L}{n+i}\C{L}{n+j}\C{L}{n-i-j}\\
F_{\ref{Binomial CC CCC+}-\ref{Binomial CC CCC-}}(i)&=\left(i+\frac{j}{2}\right)\C{L}{n-i}\C{L}{n-j}\C{L}{n+i+j}
\end{align}
and the $G$ functions are
\begin{align}
G_{\ref{Binomial kCC}}(k)&=-\frac{(L-n+k)(n+k)}{2L}\C{L}{n+k}\C{L}{n-k}\\
G_{\ref{Binomial CC CCC+}+\ref{Binomial CC CCC-}}(i)&=-\frac{(L-n+i+j)(n+i)}{2L}\C{L}{n+i}\C{L}{n+j}\C{L}{n-i-j}\\
G_{\ref{Binomial CC CCC+}-\ref{Binomial CC CCC-}}(i)&=-\frac{(L-n+i)(n+i+j)}{2L}\C{L}{n-i}\C{L}{n-j}\C{L}{n+i+j},
\end{align}
which can be checked easily, and gives us the wanted formulas. These discrete primitives can be found if $F$ is hypergeometric in the summation variable using Gosper's algorithm \cite{PWZ97.1}. We used here an implementation of Gosper's algorithm written by Peter Paule and Markus Schorn \cite{PS95.1}. If an hypergeometric ``discrete primitive'' exists, then Gosper's algorithm will find it. For the other single variable summations, the algorithm fails: there is no hypergeometric discrete primitive.\\\indent
We now move on to the last 2 simple sums (\ref{Binomial CC}) and (\ref{Binomial k2CC}). Here, the boundaries of the sums are important: with other boundaries, there might not be a simple formula for the sum. We will calculate them using a suitable generating function. Starting from the term in $t^{2n}$ in the expansion in powers of $t$ of $(1+xt)^{L}(1+yt)^{L}$, and keeping out of the sums the terms in $t^{n}$, we get
\begin{equation}
\left[\frac{(1+xt)^{L}(1+yt)^{L}}{x^{n}y^{n}}\right]_{(t^{2n})}=\C{L}{n}^{2}+\sum_{i>0}\left(\frac{x^{i}}{y^{i}}+\frac{y^{i}}{x^{i}}\right)\C{L}{n+i}\C{L}{n-i}.
\end{equation}
In $x=y=1$, it gives (\ref{Binomial CC}). If we take the derivative with respect to $x$ and to $y$ and then set $x$ and $y$ to 1, we get (\ref{Binomial k2CC}). We can now sum (\ref{Binomial CC CCC+}) and (\ref{Binomial CC CCC-}) over $j$ and use (\ref{Binomial CC}) and (\ref{Binomial kCC}) to prove (\ref{Binomial (i+j)CCC+}) and (\ref{Binomial (i+j)CCC-}). Multiplying (\ref{Binomial CC CCC+}) by $j$ and summing over $j$ gives us a linear combination of (\ref{Binomial ijCCC+}) and (\ref{Binomial (i2+j2)CCC+}), namely $(\ref{Binomial ijCCC+})+\frac{1}{4}(\ref{Binomial (i2+j2)CCC+})$.\\\indent
At this point, only (\ref{Binomial CCC+}) and either (\ref{Binomial ijCCC+}) or (\ref{Binomial (i2+j2)CCC+}) are left. We will calculate them using a generating function, as for (\ref{Binomial CC}) and (\ref{Binomial k2CC}). Starting from the term in $t^{3n}$ in the expansion in powers of $t$ of $(1+xt)^{L}(1+yt)^{L}(1+zt)^{L}$, and keeping out of the sums the terms in $t^{n}$, we get
\begin{align}
\left[\frac{(1+xt)^{L}(1+yt)^{L}(1+zt)^{L}}{x^{n}y^{n}z^{n}}\right]_{(t^{3n})}=&\C{L}{n}^{3}+\C{L}{n}\sum_{i>0}\left(\frac{x^{i}}{y^{i}}+\frac{x^{i}}{z^{i}}+\frac{y^{i}}{x^{i}}+\frac{y^{i}}{z^{i}}+\frac{z^{i}}{x^{i}}+\frac{z^{i}}{y^{i}}\right)\C{L}{n+i}\C{L}{n-i}\nonumber\\
&+\sum_{i>0}\sum_{j>0}\left[\left(\frac{x^{i}y^{j}}{z^{i+j}}+\frac{x^{i}z^{j}}{y^{i+j}}+\frac{y^{i}z^{j}}{x^{i+j}}\right)\C{L}{n+i}\C{L}{n+j}\C{L}{n-i-j}\right.\\
&\qquad\qquad\qquad\left.+\left(\frac{z^{i+j}}{x^{i}y^{j}}+\frac{y^{i+j}}{x^{i}z^{j}}+\frac{x^{i+j}}{y^{i}z^{j}}\right)\C{L}{n-i}\C{L}{n-j}\C{L}{n+i+j}\right]\nonumber.
\end{align}
In $x=y=z=1$, it gives (\ref{Binomial CCC+}). If we take the derivative with respect to $x$ and to $z$ and then set $x$, $y$ and $z$ to 1, we get another linear combination of (\ref{Binomial ijCCC+}) and (\ref{Binomial (i2+j2)CCC+}), $-(\ref{Binomial ijCCC+})-(\ref{Binomial (i2+j2)CCC+})$, proving the last two binomial formulas.
\end{section}

\begin{section}{Calculation of the third cumulant}
\label{appendix ordre3}
In this appendix, we calculate the third cumulant for systems for which $L\leq 2n$. It will be easier because all the sums will have their bounds between 0 and $2n$. In the end, using the particle-hole symmetry, we will notice that our result holds in fact for all value of $L$.

\begin{subsection}{Calculation of $g_{3}(t)$}
Noticing that for any function $f$, $[f(xt)]_{(a)}=x^{a}[f(t)]_{(a)}$, the recurrence equation (\ref{recurrg}) for $g(t)$ can be written in the more compact form
\begin{align}
&\tilde{g}_{r}(t)=-\sum_{k=1}^{r-1}\sum_{\substack{b=1\\(b\neq n)}}^{(r-k)n}\frac{L^{k}}{k!}\frac{t^{-b}}{1-x^{b-n}}\left[g_{r-k}\right]_{(-b)}\\
&+\sum_{k=1}^{r-1}\sum_{a=0}^{L}\sum_{b=1}^{(r-k)n}\sum_{c=1}^{kn}\frac{x^{c-n}\openone_{a+n\neq b+c}\openone_{a<b+c}}{(1-x^{b+c-a-n})t^{b+c-a}}(1-\openone_{a<b}-\openone_{a<c})[(1-t)^{L}]_{(a)}[g_{r-k}]_{(-b)}[g_{k}]_{(-c)}\nonumber
\end{align}
with
\begin{equation}
\openone_{u\neq v}=\left\{\begin{array}{l}1\text{ if $u\neq v$}\\0\text{ if $u=v$}\end{array}\right.\qquad\text{and}\qquad\openone_{u<v}=\left\{\begin{array}{l}1\text{ if $u<v$}\\0\text{ if $u\geq v$}\end{array}\right..
\end{equation}
Introducing the notation
\begin{equation}
I[a,b]=[(1-t)^{L}]_{(a)}[g_{2}]_{(-b)}[g_{1}]_{(-n)}=\nu(-1)^{a}\C{L}{a}[g_{2}]_{(-b)},
\end{equation}
we get at order $r=3$
\begin{equation}
\label{g3.1}
\tilde{g}_{3}(t)=-\sum_{\substack{j=1\\(j\neq n)}}^{2n}\frac{Lt^{-j}}{1-x^{j-n}}[g_{2}]_{(-j)}+\sum_{i=0}^{L}\sum_{j=1}^{2n}\openone_{i\neq j}\openone_{i<j+n}(1-\openone_{i<j}-\openone_{i<n})t^{i-j-n}\frac{1+x^{j-n}}{1-x^{j-i}}I[i,j],
\end{equation}
which will give us the solution of equation (\ref{ebQR}) corresponding to the stationary state provided (\ref{conditionrecurrg}) is satisfied, that is
\begin{equation}
\sum_{i=0}^{L}\sum_{j=1}^{2n}\delta_{ij}(1-\openone_{i<j}-\openone_{i<n})(1+x^{j-n})I[i,j]=L[g_{2}]_{(-n)}+\frac{L^{2}}{2}[g_{1}]_{(-n)},
\end{equation}
which is easily proved. From now on, we will consider that $L\leq 2n$. Then, thanks to formula (\ref{vpgtilde}), we will only need the powers of $t$ ranging from $t^{-2n}$ to $t^{-1}$ in $g(t)$ to be able to calculate the third order of the eigenvalue in the case $L\leq 2n$. We will note this $[\tilde{g}_{3}(t)]_{-2n}^{-1}$. In the double sum of equation (\ref{g3.1}), $\openone_{i<j+n}(1-\openone_{i<j}-\openone_{i<n})$ is either equal to $-1$, $0$ or $1$. The only indices $(i,j)$ for which $\openone_{i<j+n}(1-\openone_{i<j}-\openone_{i<n})$ is nonzero are (see fig. \ref{fig sums})
\begin{equation}
\label{rangesummation}
\begin{array}{ccccl}
0\leq i\leq n-1 &\quad& i+1\leq j\leq i+n &\qquad& \openone_{i<j+n}(1-\openone_{i<j}-\openone_{i<n})=-1\\
n\leq i\leq 2n &\quad& i-n+1\leq j\leq i-1 &\qquad& \openone_{i<j+n}(1-\openone_{i<j}-\openone_{i<n})=+1.
\end{array}
\end{equation}
\begin{figure}
\includegraphics[scale=0.5]{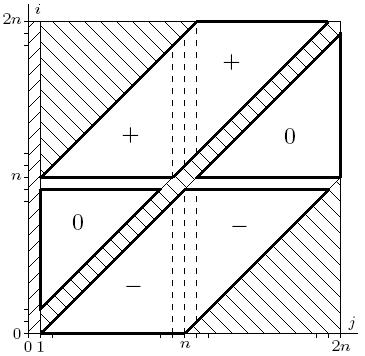}
\caption{Range of summation over $i$ and $j$ given in equation (\ref{rangesummation}). The $+$ indicate that $\openone_{i<j+n}(1-\openone_{i<j}-\openone_{i<n})=+1$, the $-$ that $\openone_{i<j+n}(1-\openone_{i<j}-\openone_{i<n})=-1$, and the $0$ that $\openone_{i<j+n}(1-\openone_{i<j}-\openone_{i<n})=0$ due to a cancellation of a $+1$ with a $-1$. The hatched parts of the graph indicate terms that do not appear in the sums, because of $\openone_{i\neq j}$, $\openone_{i<j+n}$, or because of the sums boundaries. The boundaries of the $+$, $-$ and $0$ parts of the graph belong to these parts. The boundaries of the hatched parts do not belong to them.}
\label{fig sums}
\end{figure}
giving
\begin{equation}
[\tilde{g}_{3}(t)]_{-2n}^{-1}=-L\sum_{\substack{j=1\\(j\neq n)}}^{2n}\frac{t^{-j}[g_{2}]_{(-j)}}{1-x^{j-n}}
+\sum_{i=n}^{2n}\sum_{j=i-n+1}^{i-1}\frac{1+x^{j-n}}{1-x^{j-i}}t^{i-j-n}I[i,j]-\sum_{i=0}^{n-1}\sum_{j=i+1}^{i+n}\frac{1+x^{j-n}}{1-x^{j-i}}t^{i-j-n}I[i,j].
\end{equation}
We now split the sums over $j$ according to the expression of $g_{2}(t)$
\begin{align}
[\tilde{g}_{3}(t)]_{-2n}^{-1}=&-L\sum_{j=1}^{n-1}\frac{t^{-j}[g_{2}]_{(-j)}}{1-x^{j-n}}-L\sum_{j=n+1}^{2n}\frac{t^{-j}[g_{2}]_{(-j)}}{1-x^{j-n}}+\sum_{i=n+1}^{2n-1}I[i,n]\frac{2t^{i-2n}}{1-x^{n-i}}-\sum_{i=0}^{n-1}I[i,n]\frac{2t^{i-2n}}{1-x^{n-i}}\nonumber\\
&+\sum_{i=n}^{2n}\sum_{j=i-n+1}^{n-1}I[i,j]t^{i-j-n}\frac{1+x^{j-n}}{1-x^{j-i}}+\sum_{i=n}^{2n}\sum_{j=n+1}^{i-1}I[i,j]t^{i-j-n}\frac{1+x^{j-n}}{1-x^{j-i}}\nonumber\\
&-\sum_{i=0}^{n-1}\sum_{j=i+1}^{n-1}I[i,j]t^{i-j-n}\frac{1+x^{j-n}}{1-x^{j-i}}-\sum_{i=0}^{n-1}\sum_{j=n+1}^{i+n}I[i,j]t^{i-j-n}\frac{1+x^{j-n}}{1-x^{j-i}}.
\end{align}
We switch the sums over $i$ and $j$ in the double sums and then respectively make the following changes of variables in the sums
\begin{equation}
\begin{array}{lllll}
\text{Sum 1:}\quad & j\rightarrow n-j &\qquad& \text{Sum 2:}\quad & j\rightarrow n+j\\
\text{Sum 3:}\quad & i\rightarrow n+i &\qquad& \text{Sum 4:}\quad & i\rightarrow n-i\\
\text{Sum 5:}\quad & j\rightarrow n-j\quad\text{and}\quad i\rightarrow n+i &\qquad& \text{Sum 6:}\quad & j\rightarrow n+j\quad\text{then}\quad i\rightarrow n+i+j\\
\text{Sum 7:}\quad & j\rightarrow n-j\quad\text{then}\quad i\rightarrow n-i-j &\qquad& \text{Sum 8:}\quad & j\rightarrow n+j\quad\text{and}\quad i\rightarrow n-i
\end{array}
\end{equation}
We obtain
\begin{align}
[\tilde{g}_{3}(t)]_{-2n}^{-1}=&-L\sum_{j=1}^{n-1}\frac{[g_{2}]_{(-(n-j))}}{t^{n-j}}\frac{1}{1-x^{-j}}-L\sum_{j=1}^{n}\frac{[g_{2}]_{(-(n+j))}}{t^{n+j}}\frac{1}{1-x^{j}}\\
&+\sum_{i=1}^{n-1}\frac{I[n+i,n]}{t^{n-i}}\frac{2}{1-x^{-i}}-\sum_{i=1}^{n}\frac{I[n-i,n]}{t^{n+i}}\frac{2}{1-x^{i}}\nonumber\\
&+\sum_{j=1}^{n}\sum_{i=0}^{n-1-j}\frac{I[n+i,n-j]}{t^{n-i-j}}\frac{1+x^{-j}}{1-x^{-i-j}}+\sum_{j=1}^{n}\sum_{i=1}^{n-j}\frac{I[n+i+j,n+j]}{t^{n-i}}\frac{1+x^{j}}{1-x^{-i}}\nonumber\\
&-\sum_{j=1}^{n}\sum_{i=1}^{n-j}\frac{I[n-i-j,n-j]}{t^{n+i}}\frac{1+x^{-j}}{1-x^{i}}-\sum_{j=1}^{n}\sum_{i=1}^{n-j}\frac{I[n-i,n+j]}{t^{n+i+j}}\frac{1+x^{j}}{1-x^{i+j}}\nonumber.
\end{align}
Using the expression (\ref{g2}) for $g_{2}(t)$ we can finally write, cancelling a few terms and taking the lower boundaries of all the sums equal to $1$
\begin{align}
\label{g3}
&[\tilde{g}_{3}(t)]_{-2n}^{-1}=L\nu^{2}\sum_{j=1}^{n-1}\frac{\C{L}{n+j}(-1)^{n-j}}{t^{n-j}}\frac{x^{2j}}{(1-x^{j})^{2}}+L\nu^{2}\sum_{j=1}^{n}\frac{\C{L}{n-j}(-1)^{n+j}}{t^{n+j}}\frac{x^{j}}{(1-x^{j})^{2}}\\
&-\frac{2\nu^{4}}{L}\left(\sum_{j=1}^{n}{\C{L}{n+j}\C{L}{n-j}\frac{1+x^{j}}{1-x^{j}}}\right)\left(\sum_{i=1}^{n-1}{\frac{\C{L}{n+i}(-1)^{n+i}}{t^{n-i}}\frac{x^{i}}{1-x^{i}}}+\sum_{i=1}^{n}{\frac{\C{L}{n-i}(-1)^{n-i}}{t^{n+i}}\frac{1}{1-x^{i}}}\right)\nonumber\\
&+\sum_{j=1}^{n}\sum_{i=1}^{n-1-j}\C{L}{n+i}\C{L}{n+j}\frac{\nu^{3}(-1)^{i+j}}{t^{n-i-j}}\frac{1+x^{j}}{1-x^{j}}\frac{x^{i+j}}{1-x^{i+j}}+\sum_{j=1}^{n}\sum_{i=1}^{n}\C{L}{n+i+j}\C{L}{n-j}\frac{\nu^{3}(-1)^{i}}{t^{n-i}}\frac{1+x^{j}}{1-x^{j}}\frac{x^{i}}{1-x^{i}}\nonumber\\
&+\sum_{j=1}^{n}\sum_{i=1}^{n}\C{L}{n-i-j}\C{L}{n+j}\frac{\nu^{3}(-1)^{i}}{t^{n+i}}\frac{1+x^{j}}{1-x^{j}}\frac{1}{1-x^{i}}+\sum_{j=1}^{n}\sum_{i=1}^{n-j}\C{L}{n-i}\C{L}{n-j}\frac{\nu^{3}(-1)^{i+j}}{t^{n+i+j}}\frac{1+x^{j}}{1-x^{j}}\frac{1}{1-x^{i+j}}\nonumber.
\end{align}
\end{subsection}

\begin{subsection}{Calculation of $E_{3}$}
Using (\ref{E[cumulants]}) and (\ref{vpgtilde}), we express the third cumulant as
\begin{equation}
\frac{E_{3}}{6p}=(1-x)\sum_{l>0}{\left[\C{L}{l}(-1)^{l}\frac{l(n-l)}{L(L-1)}\tilde{g}_{3}(t)\right]_{(-l)}}-(1-x)\frac{n^{3}}{3}\frac{L-n}{L-1}-n\frac{\Delta}{2p}.
\end{equation}
We insert the expression (\ref{g3}) of $\tilde{g}_{3}(t)$ in the latter equation, move the sum over $l$ inside all the other sums containing $t$ and finally set $l$ to either $n-j$, $n+j$, $n-i$, $n+i$, $n-i-j$ or $n+i+j$ according the sum using the $[\quad]_{-l}$. Rewriting
\begin{equation}
\frac{x^{k}}{1-x^{k}}=\frac{1}{2}\left(\frac{1+x^{k}}{1-x^{k}}-1\right)
\end{equation}
and
\begin{equation}
\frac{1}{1-x^{k}}=\frac{1}{2}\left(\frac{1+x^{k}}{1-x^{k}}+1\right)
\end{equation}
for all $k=i$, $j$ or $i+j$ in (\ref{g3}) and using
\begin{equation}
\frac{1+x^{i+j}}{1-x^{i+j}}\left(\frac{1+x^{i}}{1-x^{i}}+\frac{1+x^{j}}{1-x^{j}}\right)=1+\frac{1+x^{i}}{1-x^{i}}\frac{1+x^{j}}{1-x^{j}}
\end{equation}
and the binomial formulas (\ref{Binomial kCC}) (two times), (\ref{Binomial ijCCC+}), (\ref{Binomial (i2+j2)CCC+}) and (\ref{Binomial (i+j)CCC-}), which allow us to calculate the sums not involving $x$, we get
\begin{align}
&\frac{(L-1)E_{3}}{6p(1-x)}=L^{2}\sum_{i>0}\sum_{j>0}\frac{\C{L}{n+i}\C{L}{n-i}\C{L}{n+j}\C{L}{n-j}}{\C{L}{n}^{4}}(i^{2}+j^{2})\frac{1+x^{i}}{1-x^{i}}\frac{1+x^{j}}{1-x^{j}}\\
&-L^{2}\sum_{i>0}\sum_{j>0}\frac{\C{L}{n+i}\C{L}{n+j}\C{L}{n-i-j}+\C{L}{n-i}\C{L}{n-j}\C{L}{n+i+j}}{\C{L}{n}^{3}}\frac{i^{2}+ij+j^{2}}{2}\frac{1+x^{i}}{1-x^{i}}\frac{1+x^{j}}{1-x^{j}}\nonumber\\
&-L^{2}\sum_{i>0}\frac{\C{L}{n+i}\C{L}{n-i}}{\C{L}{n}^{2}}\frac{i^{2}}{2}\left(\frac{1+x^{i}}{1-x^{i}}\right)^{2}+\frac{L^{2}n(L-n)}{4(2L-1)}\frac{\C{2L}{2n}}{\C{L}{n}^{2}}-\frac{L^{2}n(L-n)}{6(3L-1)}\frac{\C{3L}{3n}}{\C{L}{n}^{3}}\nonumber\\
&+L^{2}\sum_{j>0}\frac{1+x^{j}}{1-x^{j}}\left(\frac{\C{L}{n+j}\C{L}{n-j}}{\C{L}{n}^{2}}\frac{j^{2}(L-2n)}{2L}\right.\nonumber\\
&\qquad\qquad\qquad\qquad\left.+\sum_{i>0}\frac{\C{L}{n+i}\C{L}{n+j}\C{L}{n-i-j}-\C{L}{n-i}\C{L}{n-j}\C{L}{n+i+j}}{\C{L}{n}^{3}}\left(\frac{j^{2}}{2}+ij\right)\right)\nonumber\\
&+nL^{2}\sum_{j>0}\frac{1+x^{j}}{1-x^{j}}\left(\frac{\C{L}{n+j}\C{L}{n-j}}{\C{L}{n}^{2}}\left(\frac{n(L-n)}{L}-\frac{j}{2}\right)\right.\nonumber\\
&\qquad\qquad\qquad\qquad\qquad\left.-\sum_{i>0}\frac{\C{L}{n+i}\C{L}{n+j}\C{L}{n-i-j}+\C{L}{n-i}\C{L}{n-j}\C{L}{n+i+j}}{\C{L}{n}^{3}}\left(\frac{j}{2}+i\right)\right)\nonumber.
\end{align}
The last two sums identically vanish thanks to the binomial formulas (\ref{Binomial CC CCC+}) and (\ref{Binomial CC CCC-}). This ends the proof of formula (\ref{cumulant3}) for the third cumulant.
\end{subsection}
\end{section}

\begin{section}{Subleading corrections to the third cumulant in the scaling limit}
\label{appendix Phi0}
In this appendix, we compute the two first subleading corrections to the integral formula (\ref{cumulant3 integral}) for the third cumulant starting from the finite size expression (\ref{cumulant3}). We will need subleading corrections to the binomial coefficient formulas that we used to derive the integral expression (\ref{cumulant3 integral}). In particular, we will use
\begin{equation}
\frac{\C{kL}{kn}}{\C{L}{n}^{k}}=\sqrt{\frac{(2\pi\rho(1-\rho)L)^{k-1}}{k}}+\frac{(1-\rho+\rho^{2})(k^{2}-1)}{12}\sqrt{\frac{(2\pi)^{k-1}(\rho(1-\rho)L)^{k-3}}{k^{3}}}+\O{L^{\frac{k-5}{2}}}
\end{equation}
and
\begin{align}
\frac{\C{L}{n+i}}{\C{L}{n}}=\left(\frac{1-\rho}{\rho}\right)^{i}e^{-\frac{i^{2}+(1-2\rho)i}{2\rho(1-\rho)L}}&\left(1+\frac{1}{L^{2}}\left(\frac{(1-2\rho)i^{3}}{6\rho^{2}(1-\rho)^{2}}+\frac{(1-2\rho+2\rho^{2})i^{2}}{4\rho^{2}(1-\rho)^{2}}+\cdots\right)\right.\\
&\qquad\left.+\frac{1}{L^{3}}\left(-\frac{(1-3\rho+3\rho^{2})i^{4}}{12\rho^{3}(1-\rho)^{3}}+\cdots\right)+\frac{1}{L^{4}}\left(\frac{(1-2\rho)^{2}i^{6}}{72\rho^{4}(1-\rho)^{4}}+\cdots\right)+\O{\frac{1}{L^{5}}}\right)\nonumber
\end{align}
The $\cdots$ terms contain lower powers in $i$ that come from subdominant corrections to the Stirling formula. They will not contribute to the corrections of $E_{3}$ we will calculate. In the following, we will also need an expansion of the difference between a Riemann sum and the corresponding limit integral. Let $f$ be a function of $2$ variables with a finite limit when any of the two variables goes to $0$ and decaying exponentially at infinity. We want to calculate the difference between the sum
\begin{equation}
S=\epsilon^{2}\sum_{i=1}^{\infty}\sum_{j=1}^{\infty}f(i,j)
\end{equation}
and the integral
\begin{equation}
I=\int_{0}^{\infty}\!\int_{0}^{\infty}dudvf(u,v)
\end{equation}
at order $2$ in $\epsilon$. Writing $I$ as
\begin{align}
I=\sum_{i=0}^{\infty}\sum_{j=0}^{\infty}\int_{i\epsilon}^{(i+1)\epsilon}\!\int_{j\epsilon}^{(j+1)\epsilon}dudv&\left(f(i\epsilon,j\epsilon)+(u-i\epsilon)\partial_{u}f(i\epsilon,j\epsilon)+(v-j\epsilon)\partial_{v}f(i\epsilon,j\epsilon)+\frac{(u-i\epsilon)^{2}}{2}\partial_{u}^{2}f(i\epsilon,j\epsilon)\right.\nonumber\\
&\qquad\qquad\left.+(u-i\epsilon)(v-j\epsilon)\partial_{u}\partial_{v}f(i\epsilon,j\epsilon)+\frac{(v-j\epsilon)^{2}}{2}\partial_{v}^{2}f(i\epsilon,j\epsilon)+\cdots\right)
\end{align}
and taking care of the fact that the sums begin with $i$ and $j$ equal to $1$, we find
\begin{equation}
S=I-\frac{\epsilon}{2}\left(\int_{0}^{\infty}duf(u,0)+\int_{0}^{\infty}dvf(0,v)\right)-\epsilon^{2}\left(-\frac{f(0,0)}{4}+\frac{1}{12}\int_{0}^{\infty}du\partial_{v}f(u,0)+\frac{1}{12}\int_{0}^{\infty}dv\partial_{u}f(0,v)\right)+\O{\epsilon^{2}}
\end{equation}
Using the finite size equation (\ref{cumulant3}), we now can calculate subleading corrections to $E_{3}/(1-x)$. The term of order $L^{3}$ we get is the one we found previously (\ref{cumulant3 scaling}). At the next order, we find that the $L^{5/2}$ term cancels out. Finally, the term of order $L^{2}$ is given by
\begin{align}
\label{cumulant3 subleadingintegral}
\left[\frac{E_{3}}{1-x}\right]_{(L^{2})}=&6\rho^{2}(1-\rho)^{2}L^{2}\int_{0}^{\infty}\!\int_{0}^{\infty}dudv\frac{(u^{2}+v^{2})e^{-u^{2}-v^{2}}-(u^{2}+uv+v^{2})e^{-u^{2}-uv-v^{2}}}{\tanh(\Phi u)\tanh(\Phi v)}\nonumber\\
&+3\rho(1-\rho)(1-2\rho+2\rho^{2})L^{2}\int_{0}^{\infty}\!\int_{0}^{\infty}dudv\frac{(u^{2}+v^{2})^{2}e^{-u^{2}-v^{2}}-(u^{2}+uv+v^{2})^{2}e^{-u^{2}-uv-v^{2}}}{\tanh(\Phi u)\tanh(\Phi v)}\nonumber\\
&-\rho(1-\rho)(1-3\rho+3\rho^{2})L^{2}\int_{0}^{\infty}\!\int_{0}^{\infty}dudv\frac{(u^{2}+v^{2})(u^{4}+v^{4})e^{-u^{2}-v^{2}}-(u^{2}+uv+v^{2})^{3}e^{-u^{2}-uv-v^{2}}}{\tanh(\Phi u)\tanh(\Phi v)}\nonumber\\
&+\frac{3}{4}\rho(1-\rho)(1-2\rho)^{2}L^{2}\int_{0}^{\infty}\!\int_{0}^{\infty}dudv\frac{-u^{2}v^{2}(u+v)^{2}(u^{2}+uv+v^{2})e^{-u^{2}-uv-v^{2}}}{\tanh(\Phi u)\tanh(\Phi v)}\nonumber\\
&-\frac{20\pi}{27\sqrt{3}}\rho^{2}(1-\rho)^{2}L^{2}-\frac{4\pi}{27\sqrt{3}}\rho(1-\rho)L^{2}-\rho(1-\rho)L^{2}\int_{0}^{\infty}\frac{\left(u^{2}-\frac{u^{4}}{2}\right)e^{-u^{2}}}{\Phi\tanh(\Phi u)}
\end{align}
This expression can be easily checked numerically: subtracting this expression plus the $L^{3}$ leading term to the finite size expression (\ref{cumulant3}) for various values of $\Phi$, $n$ and $L$, we find that the remaining term is of order $L\ll L^{2}$. Taking now the limit $\Phi\to 0$ in expression (\ref{cumulant3 subleadingintegral}), we find that the divergent term vanishes as expected, leaving us with
\begin{equation}
\left[\frac{E_{3}}{1-x}\right]_{(L^{2})}\to\rho^{2}(1-\rho)^{2}L^{2}\quad\text{when $\Phi\to 0$}
\end{equation}
\end{section}

\bibliographystyle{unsrt}
\bibliography{}

\end{document}